\documentclass[11pt, preprint]{aastex}
\usepackage{graphicx}
\usepackage{natbib}
\usepackage{float}
\pdfoutput=1
\shorttitle{A New Deeply-Eclipsing Intermediate Polar}
\begin{document}
\title{CXOGBS J174954.5-294335: A New Deeply-Eclipsing Intermediate Polar}
\author{\bf{Christopher B. Johnson}}
\affil{Department of Physics and Astronomy, Louisiana State University, Baton Rouge, LA 70803-4001, USA}
\affil{Space Telescope Science Institute, 3700 San Martin Drive, Baltimore, MD, 21218, USA}
\author{Manuel A. P. Torres}
\affil{SRON, Netherlands Institute for Space Research, Sorbonnelaan 2, 3584 CA, Utrecht, The Netherlands}
\affil{Department of Astrophysics/IMAPP, Radboud University Nijmegen, Heyendaalseweg 135, 6525 AJ, Nijmegen, The Netherlands}
\author{Robert I. Hynes}
\affil{Department of Physics and Astronomy, Louisiana State University, Baton Rouge, LA 70803-4001, USA}
\author{Peter G. Jonker}
\affil{SRON, Netherlands Institute for Space Research, Sorbonnelaan 2, 3584 CA, Utrecht, The Netherlands}
\affil{Department of Astrophysics/IMAPP, Radboud University Nijmegen, Heyendaalseweg 135, 6525 AJ, Nijmegen, The Netherlands}
\author{Craig Heinke}
\affil{Dept. of Physics, University of Alberta, Rm. 238 CEB, Edmonton, AB T6G 2G7, Canada}
\author{Thomas Maccarone}
\affil{Department of Physics and Astronomy, Texas Tech University, Box 41051, Lubbock, TX 79409-1051, USA}
\author{Christopher T. Britt}
\affil{Department of Physics and Astronomy, Texas Tech University, Box 41051, Lubbock, TX 79409-1051, USA}
\affil{Department of Physics and Astronomy, Michigan State University, 5678 Wilson Road, Lansing, MI 48824, USA}
\author{Danny Steeghs}
\affil{Department of Physics, University of Warwick, Coventry CV4 7AL, UK}
\author{Thomas Wevers}
\affil{Department of Astrophysics/IMAPP, Radboud University Nijmegen, Heyendaalseweg 135, 6525 AJ, Nijmegen, The Netherlands}
\and
\author{Jianfeng Wu}
\affil{Department of Astronomy, University of Michigan, 1085 S University Ave., Ann Arbor, MI 48109, USA}

%\email{chjohnson@stsci.edu (CBJ)}
%\date{Accepted; Received; in original form}

%\pagerange{\pageref{firstpage}--\pageref{lastpage}} \pubyear{2013}

%\maketitle

%\label{firstpage}

\begin{abstract} 
We present the results of a photometric and spectroscopic analysis of the Galactic Bulge Survey X-ray source CXOGBS J174954.5-294335 (hereafter, referred to as CX19). CX19 is a long period, eclipsing intermediate polar type cataclysmic variable with broad, single-peaked Balmer and Paschen emission lines along with He{\sc ii} $\lambda4686$ and Bowen blend emission features. With coverage of one full and two partial eclipses and archival photometry, we determine the ephemeris for CX19 to be HJD(eclipse) = 2455691.8581(5) + 0.358704(2)$\times$N. We also recovered the white dwarf spin period of P$_{\rm spin}$ = 503.32(3) seconds which gives a P$_{\rm spin}$/P$_{\rm orb}$ = 0.016(6), comparable to several confirmed, long period intermediate polars. CX19 also shows a clear X-ray eclipse in the 0.3--8.0 keV range observed with Chandra. Two optical outbursts were observed lasting between 6--8 hours (lower limits) reaching $\sim$1.3 mags in amplitude. The outbursts, both in duration and magnitude, the accretion disc dominated spectra and hard X-ray emission are reminiscent of the intermediate polar V1223 Sgr sharing many of the same characteristics. If we assume a main sequence companion, we estimate the donor to be an early G-type star and find a minimum distance of $d \approx$ 2.1 kpc and a 0.5--10.0 keV X-ray luminosity upper limit of 2.0 $\times$ 10$^{33}$ erg s$^{-1}$. Such an X-ray luminosity is consistent with a white dwarf accretor in a magnetic cataclysmic variable system. To date, CX19 is only the second deeply-eclipsing intermediate polar with X-ray eclipses and the first which is optically accessible.
 \end{abstract}

\section{Introduction} Intermediate polars (IPs) form a sub-class of cataclysmic variables (CVs) in which a magnetic white dwarf (WD) accretes matter from a late-type companion that is filling its Roche lobe. The WD's strong magnetic field truncates the accretion disc towards inner disc radii and funnels the material along the field lines causing accretion columns to form at the poles of the WD surface. Polars differ from IPs due to the intense B field ($>$10$^7$ Gauss) that has caused the WD rotation to be phase-locked to the orbital motion of the binary system (i.e. P$_{\rm spin}$ = P$_{\rm orb}$). In an IP system, the WD rotation period is much shorter than the orbital period. IPs have a well known relation between the WD spin period and the orbital period of the system with the majority of confirmed systems showing P$_{\rm spin}$/P$_{\rm orb}$ $\le$ 0.1 \citep{Norton2004}.\\
\indent The optical spectrum of IPs is usually characterized by a blue continuum and strong Balmer and He{\sc ii} $\lambda 4686$ lines originated in the cooling of shock-heated gas near the WD surface as well as in the irradiated accretion flow and donor star. At some point in the disc, the kinetic energy density of the gas is overtaken by the magnetic energy density and material will be guided along the magnetic field lines leading to radial accretion onto the poles. The release of energy in the form of accretion luminosity creates a bright spot above the WD's poles. If the WD spin and magnetic axis are not synchronized, this scenario can produce modulations in the X-ray or optical light curves at the WD spin period \citep{Patterson1994}. A fraction of the energy released from the funneled material at the poles can fall upon structures in orbit around the WD and results in a beat frequency. This frequency occurs at lower frequencies than the WD spin frequency. The frequency of this reprocessed light is known as the ``orbital sideband'' frequency and is found from the relation: $\omega_{\rm repro}$ = $\omega_{\rm spin}$ - $\omega_{\rm orb}$, although sidebands can be produced as multiples of the orbital and spin frequencies as well.\\ 
\indent A substantial portion of the energy emitted in IPs is produced in a shock above both the magnetic poles, which is optically thin even for high mass transfer rates. This is in contrast to the boundary layer in novalike CVs (NLs) or dwarf novae (DNe) in outburst which becomes optically thick and shifts most of the emitted radiation towards the ultraviolet \citep{Patterson1994}. Therefore, X-ray surveys are ideal and well-suited for the detection of IPs (and Polars).\\ 
%The best way for detecting these systems is using X-ray surveys to gather the X-ray source coordinates, accompanied by ground-based follow-up campaigns of the optical counterpart (if present).\\
\indent The Galactic Bulge Survey (GBS; \citealt{Jonker2011}, \citealt{Jonker2014}) is a wide and shallow X-ray survey of two strips both above and below the Galactic Bulge aimed at uncovering quiescent X-ray binaries harboring black holes or neutron stars. The GBS is also expected to discover a large number of CVs ($\sim$210, \citealt{Jonker2011}) in both the X-ray and optical, most of which should be IPs. Here we present the identification and analysis of an eclipsing IP in the Galactic Bulge Survey. During the ongoing Galactic Bulge Survey, we have identified several IP candidates (\citealt{Britt2013, Torres2014}) and this work presents the first in-depth, photometric and spectroscopic analysis on one source, CX19.\\ %We report the discovery of an intermediate polar found in the Galactic Bulge as an X-ray selected target during the Galactic Bulge Survey.\\ 
%\indent The first GBS X-ray source CXOGBS J174447.5-270101 (CX73, 23 X-ray counts in 2ks exposure) is an ellipsoidal, interacting binary with an orbital period of P$_{\rm orb}$ = 0.457844(5) days. The WD spin period of P$_spin$ = 0.047307(5) days and two orbital sideband periods are present in the analysis of the periodograms. CX73 has an i' magnitude of The optical spectroscopy only shows narrow, H$\alpha$ emission and no apparent companion absorption features.\\
%\indent The second GBS X-ray source CXOGBS J174954.5-294335 (CX19, 62 X-ray counts in 2ks exposure) is an eclipsing, interacting binary with a orbital period of P$_{\rm orb}$ = 0.358705(5) days. The optical spectroscopy shows strong, narrow Balmer and HeII 4686 emission in the spectra typical of a disc-accretion scenario. CX19 has an r' magnitude of 18.65(5) and was also detected in the Optical Gravitational Lensing Experiment IV (OGLE-IV) with an I magnitude of 17.91.\\
%as source number: blg501.24.34934. The OGLE I-band data set contains several years worth of data showing two $\sim$1 magnitude outbursts in the photometric light curve.\\
%\indent We begin with a short introduction and then section 2 describes the astrometry, optical, spectroscopic, and X-ray observations. Section 3 presents the results and data analysis. We end with a discussion on the nature of CX19 followed by the conclusion.\\

\section{Observations}
\subsection{4.0 meter Blanco Telescope Data} Two nights of photometry were obtained at CTIO from June 10--11, 2013 using the DECam instrument \citep{Abbott2012}. We acquired 217 data points in the SDSS r' band using 90 second exposures with seeing in the range of 0.89--1.51 arcseconds. DECam uses a 62 CCD camera with 2048$\times$4096 pixels per chip, pixel scale of 0.27 arcsec/pixel and a 2.2 square degree field of view\footnote[1]{http://www.ast.noao.edu/sites/default/files/NOAO\_DHB\_v2.2.pdf}. The DECam data were reduced with the NOAO DECam pipeline and were acquired through the NOAO science archive providing an astrometric solution accompanied by a World Coordinate System (WCS) for each image. The images were calibrated using the American Association of Variable Star Observers (AAVSO) Photometric All-Sky Survey DR9 (APASS) and carry an average uncertainty of 0.05 magnitudes. We used between 3-5 comparison stars in the field of view with uncertainties on the order of 0.001 magnitudes and propagated the error for each data point.\\ 

\subsection{Optical Gravitational Lensing Experiment Data} Archival I-band photometry was obtained from the Optical Gravitational Lensing Experiment (OGLE-IV; \citealt{Udalski2012}) online database. OGLE is an all-sky monitoring program aimed at searching for dark matter with microlensing phenomena. The I-band data were collected from the 1.3 meter Warsaw Telescope at Las Campa\~{n}as Observatory in Chile. The CCD has 32 thin, $2048\times4096$ pixel chips with a 0.26 arcsec/pixel scale and 1.4 square degree field of view. The exposure time is 150 seconds with an average time between exposures of 23 minutes. We have gathered 5252 data points from OGLE-IV fields that span from March of 2010 to July of 2012 with the field and source number: blg501.24.34934. The data were part of the original OGLE monitoring program.\\

\subsection{Very Large Telescope Spectroscopy} Spectroscopy of CX19 was obtained with the Visible Multi-Object Spectrograph (VIMOS; \citealt{Lefevre2003}) which is mounted on the Nasmyth focus of the 8.2-m European Southern Observatory (ESO) Unit 3 Very Large Telescope (VLT)  at Paranal, Chile.  The observations were acquired in service mode under program 091.D-0062(A) on July 13, 2013.  VIMOS was operated with the medium resolution (MR) grism which, in combination with the unbinned $2048\times4096$ pixel EEV CCD detectors, provided  a 2.5 \AA/pixel dispersion and a wavelength coverage of $\sim 4800-10000$ \AA. The use of 1.0'' width slits secured a spectral resolution of $\sim10$ \AA~full-width at half maximum  (FWHM). A one hour observing block (ID 975525) was executed for the spectroscopic identification of the optical counterpart to CX19 and another GBS source in the VIMOS field of view. The observing block  consisted of acquisition imaging, two spectroscopic integrations of 875 s, three flat-field exposures and a comparison lamp exposure for wavelength calibration. The standard data reduction steps were performed with the ESO-VIMOS pipeline. The {\sc iraf kpnoslit} package was used to interactively extract the two spectra from the reduced 2-D frames (see \cite{Torres2014} for further details on the data extraction). From the width of the source's spatial profile at spectral positions covering H$\alpha$, we measure an image quality of 0.8". Therefore, the observations were obtained in seeing-limited conditions yielding a spectral resolution of 8 \AA~FWHM (365 km s$^{-1}$ at H$\alpha$). The wavelength calibration zero-point was corrected by applying to the VIMOS spectra a --30 km s$^{-1}$ shift calculated using the [O{\sc i}] $\lambda6300.3$ sky line.\\ 
\indent To better constrain the nature of CX19, we obtained five 1200s spectra (taken non-sequentially to achieve as much phase coverage as possible) with the FOcal Reducer and low dispersion Spectrograph 2 (FORS2, \citealt{Appenzeller1998}) mounted on the Cassegrain focus of the 8.2-m ESO Unit 1 VLT.  The observations were obtained in visitor mode under program 093.D-0939(A) on 8 May 2014. FORS2 was used with the standard resolution collimator and the $2048\times4096$ pixel MIT  detector binned $2 \times 2$. The instrument was operated with a 0.7'' width long-slit and the GRIS\_1200g grism. The instrumental setup and object location along the slit yielded a dispersion of 0.72 \AA~ pixel$^{-1}$, a coverage in the spectral range $\lambda\lambda4090-5570$ and a slit-width limited resolution of $\sim1.4$ \AA~ FWHM. Bias, flat-field exposures and exposures using comparison arc lamps were performed after the end of the night. The spectra were reduced and extracted using standard techniques implemented in the {\sc starlink}, {\sc figaro} and {\sc pamela} packages while the wavelength calibration was done with {\sc molly}. The data reduction consisted of de-biasing and flat-fielding the data. 
The wavelength calibration was derived through polynomial fits to 13 arc lines. The root mean square (rms) scatter of the fit was $< 0.02$ \AA. Due to the lack of sky emission lines in the observed spectral range, it was not possible to evaluate the deviations in the  wavelength calibration zero-point which potentially can reach $\sim 30$ km s$^{-1}$ \citep{Torres2015}. We measure from the width of the source spatial profile at He{\sc ii} $\lambda4686$ a 0.6'' $\pm$ 0.1'' FWHM image quality. The observations yielded a spectral resolution slightly better than 90 km s$^{-1}$ FWHM at He{\sc ii} $\lambda4686$.\\
\indent Finally, the spectra were normalized by dividing each of them by the result of fitting a low-order spline function to the continuum. We masked out emission lines, diffuse interstellar bands (DIBs) and, in the case of the VIMOS data, the atmospheric absorption bands. The resulting normalized spectra were then re-binned to a uniform pixel scale.
%\indent The resulting spectra were normalized by dividing each of them by a low-order spline fit to the continuum after masking out emission lines, atmospheric absorption bands and diffuse interstellar bands. Finally, the resulting spectra were re-binned to a uniform pixel scale.

\subsection{Chandra X-ray Observatory Data} The Chandra X-ray Observatory (CXO) observations were acquired using the I0 to I3 CCDs of the Advanced CCD Imaging Spectrometer (ACIS-I) detector (\citealt{Garmire1997}; ACIS-I). The observation identification numbers (ID) for the data presented here are 8742, 8753 and 9563 and had exposure times lasting 2 ks, 2 ks and 15 ks, respectively. The data were reprocessed and analyzed using the CIAO 4.7 software developed by the Chandra X-ray Center and employs CALDB version 4.6.7. The data telemetry mode was set to `very faint' for all observations where the `very faint' mode provides 5$\times$5 pixel information per X-ray event. In our analysis, we selected events that fell in the energy range 0.3--8.0 keV. %The X-ray light curves for both sources are plotted in bins of 100 seconds to show maximum variability. 

\section{Results}
\subsection{Astrometry} The 2 ks GBS CXO pointing (OBSID 8742) provided a 95\% confidence region with a 0.89 arcsec error circle calculated using equation (5) from \cite{Hong2005}. This left two variables within the error circle region, an eclipsing binary and a red giant, both of which have data in the OGLE database. We confirmed the eclipsing binary as the optical counterpart to our X-ray coordinates by using the Chandra Source Catalog (CSC) (version 1.1) \citep{Evans2010}. A 15 ks CSC CXO pointing (OBSID 9563) provided a 95\% confidence region centered on the coordinates of the eclipsing binary with a 0.39 arcsec error circle. The DECam finder chart can be seen in Figure~1.  
\begin{figure}
\centering
\includegraphics[scale=0.65]{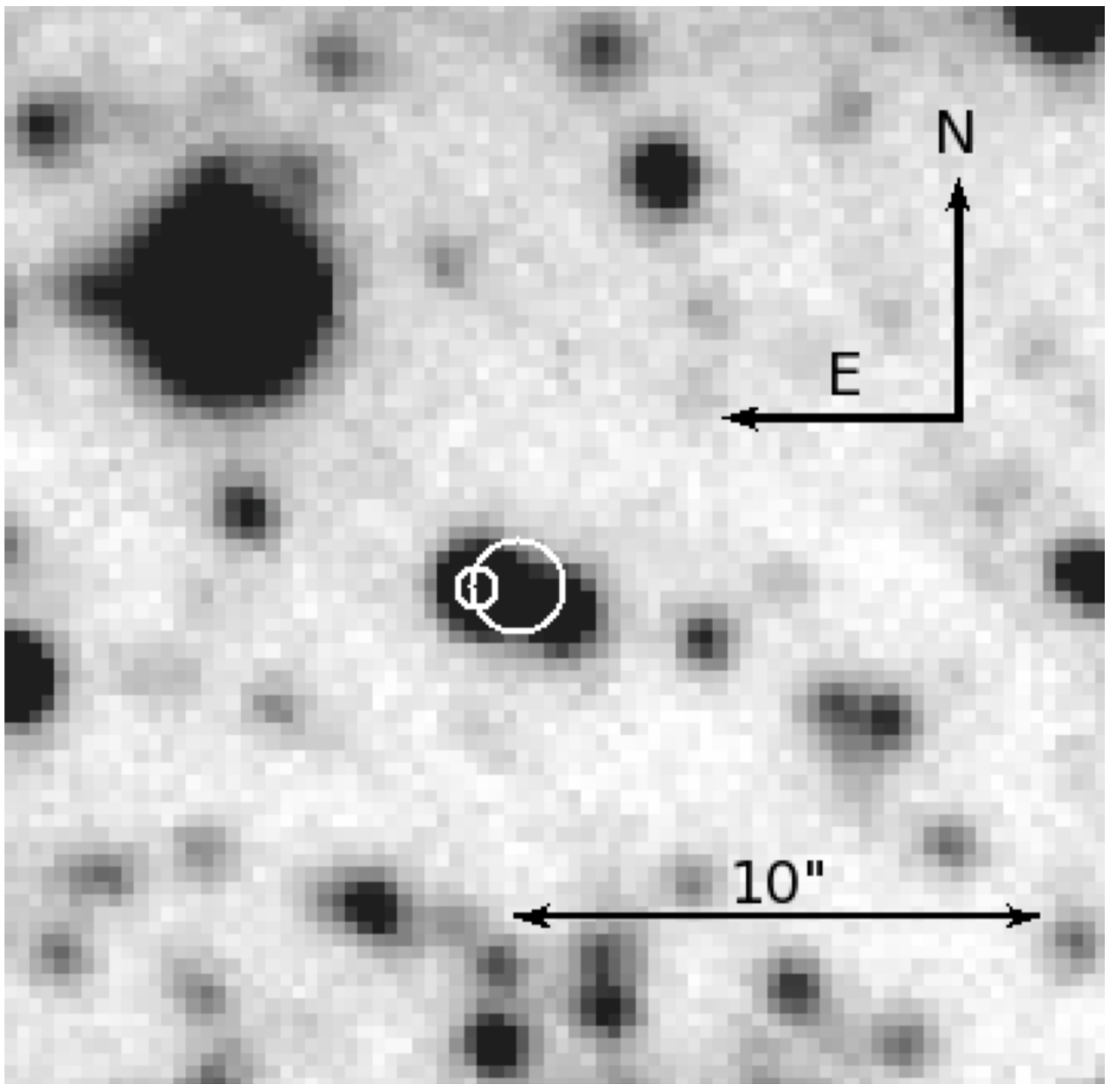}
\caption{DECam finder chart for CX19. The white circles represent 95\% confidence region error circles:  the larger circle is the GBS X-ray position and the smaller is the Chandra Source Catalog X-ray position. Two stars are consistent with the larger error circle. an eclipsing binary (East) and a red giant (West).}
\end{figure}

\subsection{Photometric Analysis} CX19 was first classified as an eclipsing source in \cite{Britt2014} based solely on data acquired with the Mosaic-II instrument. The original data only contained one eclipse point out of 20 over eight nights of observations so no orbital period could be discerned. Using DECam, we uncovered two eclipses and the egress of another eclipse over two nights. We performed an initial period search on the DECam data using the software package PERIOD\footnote[2]{ http://www.starlink.rl.ac.uk/docs/sun167.htx/sun167.html}. PERIOD is a time-series analysis package designed to search for periodicities in data sets using different analysis techniques.\\
\indent The Lomb-Scargle periodogram of the DECam data is shown in Figure~2 a). The DECam data shows a fundamental orbital frequency f$_{\rm orb}$ = 2.7878(3) cycles/day. This frequency corresponds to the orbital period of P$_{\rm orb}$ = 0.358(3) days and is consistent with the reported OGLE period (P$_{\rm rep}$ = 0.3587 days) from \cite{Udalski2012}. A second fundamental frequency (P$_{\rm spin}$) was found which we interpret as the WD spin frequency of 171.2915(9) cycles/day or P$_{\rm spin}$ = 0.005838(5) days = 504.4(5) secs.\\
\indent The DECam light curve and phase-folded light curves can be seen in Figure~3. The middle panel of Figure~3 shows the light curve folded on P$_{\rm orb}$ and binned in phase bins of three consecutive data points with a mean out-of-eclipse r' magnitude of 18.67(8), an eclipse depth of 0.53 mags and eclipse duration lasting $\sim$53 minutes. After binning the data, we find evidence of a secondary eclipse around phase 0.5 with an approximate depth of 0.11 mag. The bottom panel of Figure~3 shows a clear modulation when folded on the P$_{\rm spin}$ with amplitude of $A$ = 0.2 mag. We point out that upon inspecting the periodogram around the P$_{\rm spin}$, we do not find any orbital sideband frequencies.\\%, but frequencies that correspond to aliases.\\
%\indent The error on the periodicities (along with the rest of the errors in this paper, unless otherwise noted) were estimated by using a bootstrapping technique with a Monte Carlo approach much like \cite{tsteegh07}. The target data set was recopied, where the input data set is resampled by randomly selecting data points. The data points are allowed to be resampled more than once during the sampling and then each data set ÔcopyÕ was subjected to fitting an appropriate function to the data and allowing the parameters to vary. For each bootstrap copy, the minimum $\chi^2$ was calculated and the best fit parameters were found. Besides providing the best-fit parameter, the 1$\sigma$ error for each parameter is calculated.
\indent The errors calculated in this manuscript were estimated using a bootstrapping technique with a Monte Carlo approach. See \cite{Steeghs2007} or \cite{Johnson2014} for  a full description of the technique.

\begin{figure}
\centering
\includegraphics[scale=0.5]{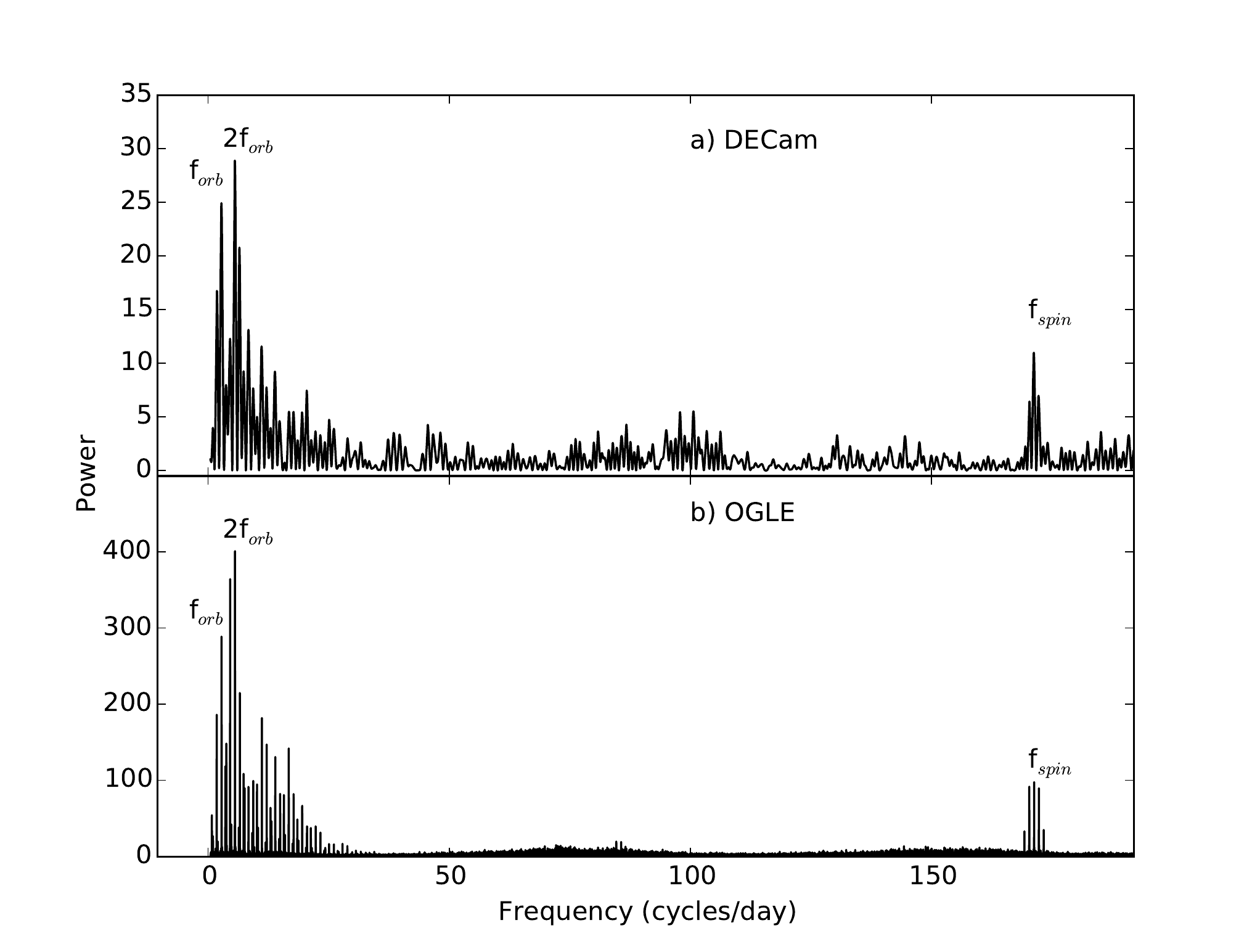}
\caption{The initial Lomb-Scargle periodograms from the a) DECam data and the b) OGLE data. The orbital and spin frequencies are labelled along with the second orbital harmonic (2f$_{\rm orb}$) as the most prominent peak.}
\end{figure}

\indent We derive times of mid-eclipse by fitting a parabolic function to the three eclipse profiles observed during the two nights of DECam observations. The parameters were allowed to vary while the $\chi^2$ was calculated for every distinct set of parameters. In parameter space, we allowed the date and magnitude of the parabola to vary in order to find the best fit for the time of minimum of the eclipse using a parabolic function of the form:
\begin{equation}
y = a\times({\rm T}_{\rm obs} - h)^2 + k
\end{equation}
where $a$ is th.pdfe width of the parabola (held constant based on the two full eclipses), $h$ and $k$ are initial guesses for the HJD and magnitude and T$_{\rm obs}$ is the observed HJD for each data point. Our best fits yield the following times of mid-eclipse: HJD = 2456454.8144(5), 2456455.5331(5) and 2456455.8940(5).\\
%\begin{equation}
%HJD(eclipse) = 2456454.81584(6) + 0.358705(5)\times N 
%\%end{equation}
%where $N$ is the cycle number. Equation (1) provided our best fit with a $\chi^2_{red}$ = 1.64 (12 degrees of freedom). %A summary of the DECam light curves can be seen in Figure~1.
\begin{figure}
\centering
\includegraphics[scale=0.43]{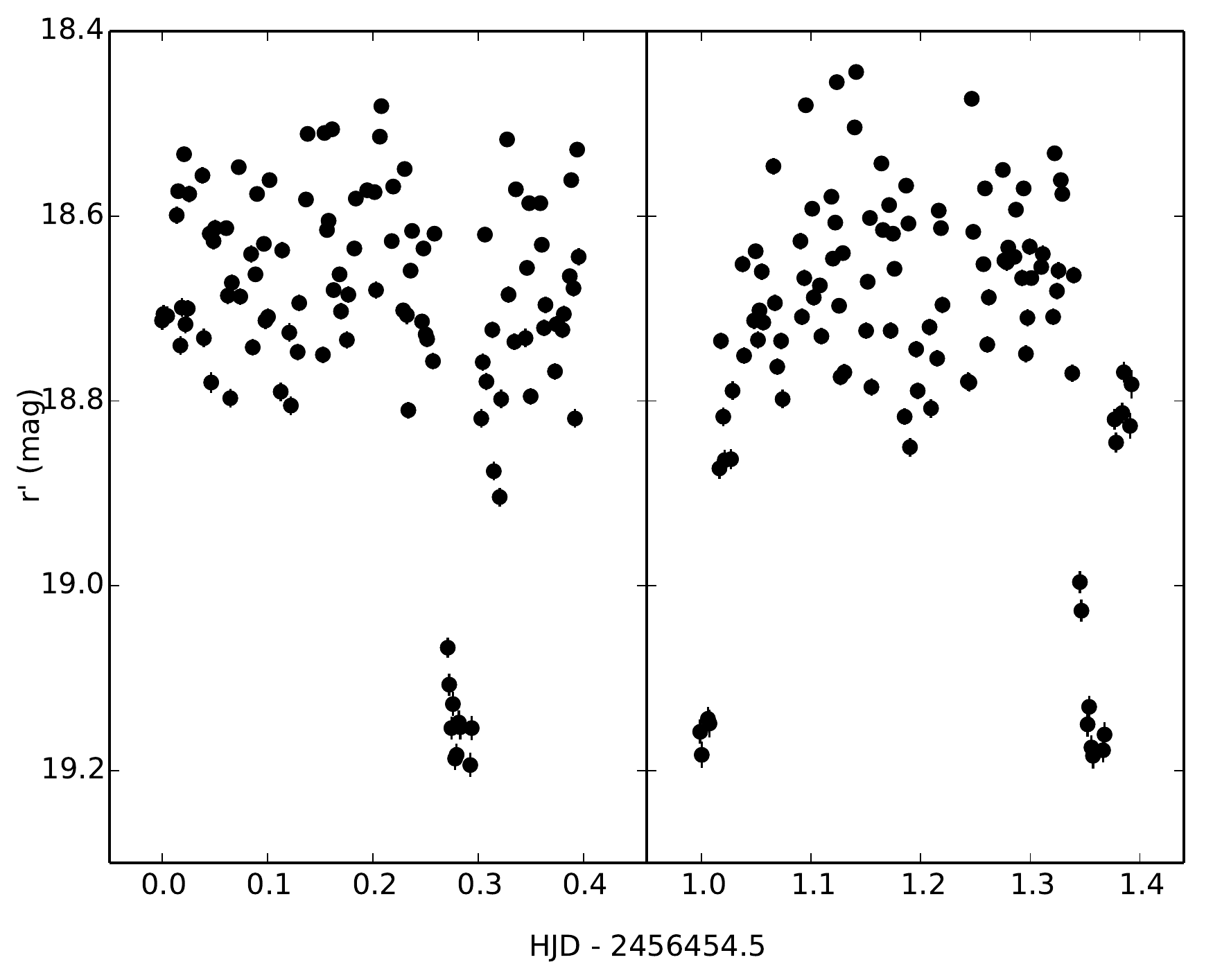}
\includegraphics[scale=0.43]{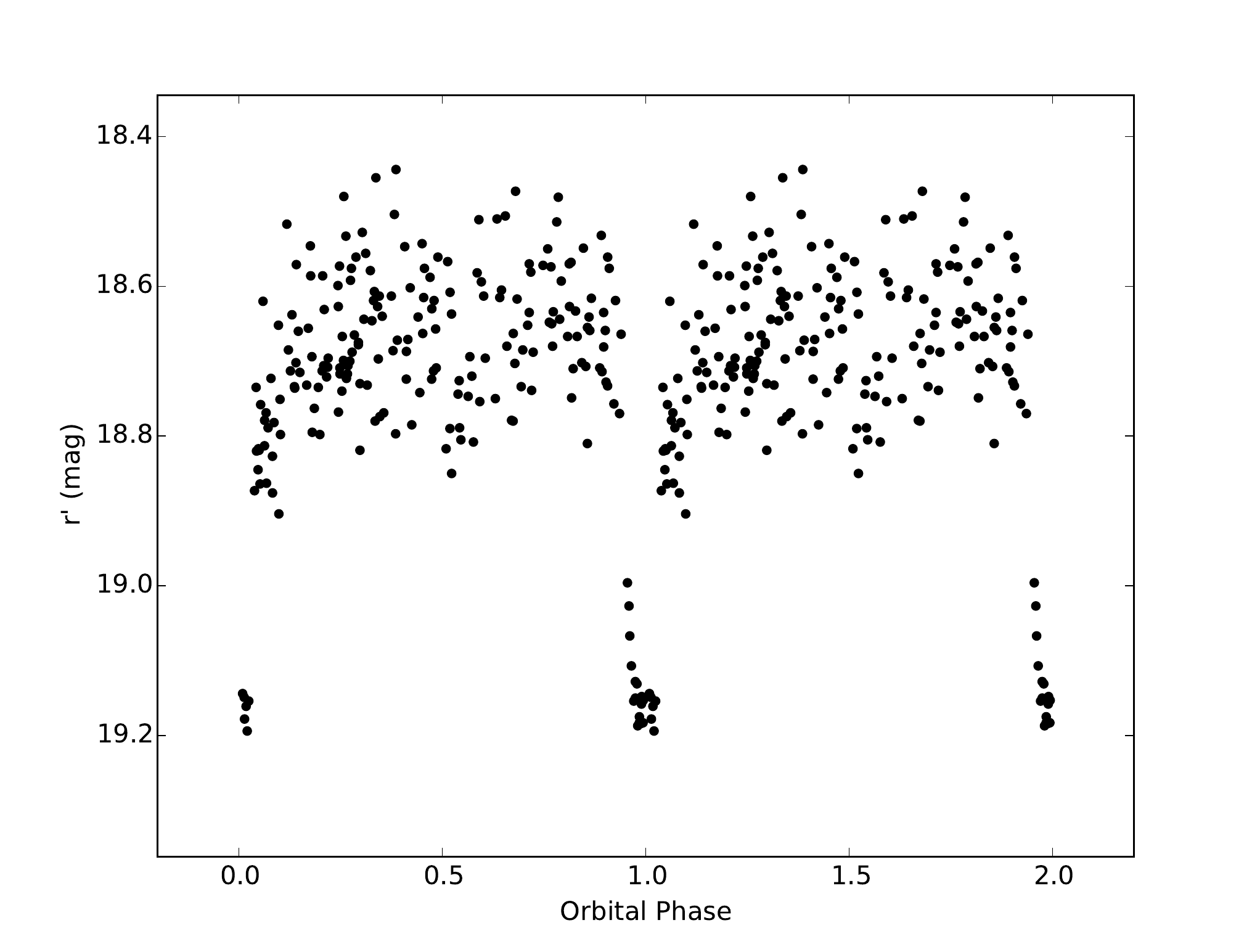}
\includegraphics[scale=0.43]{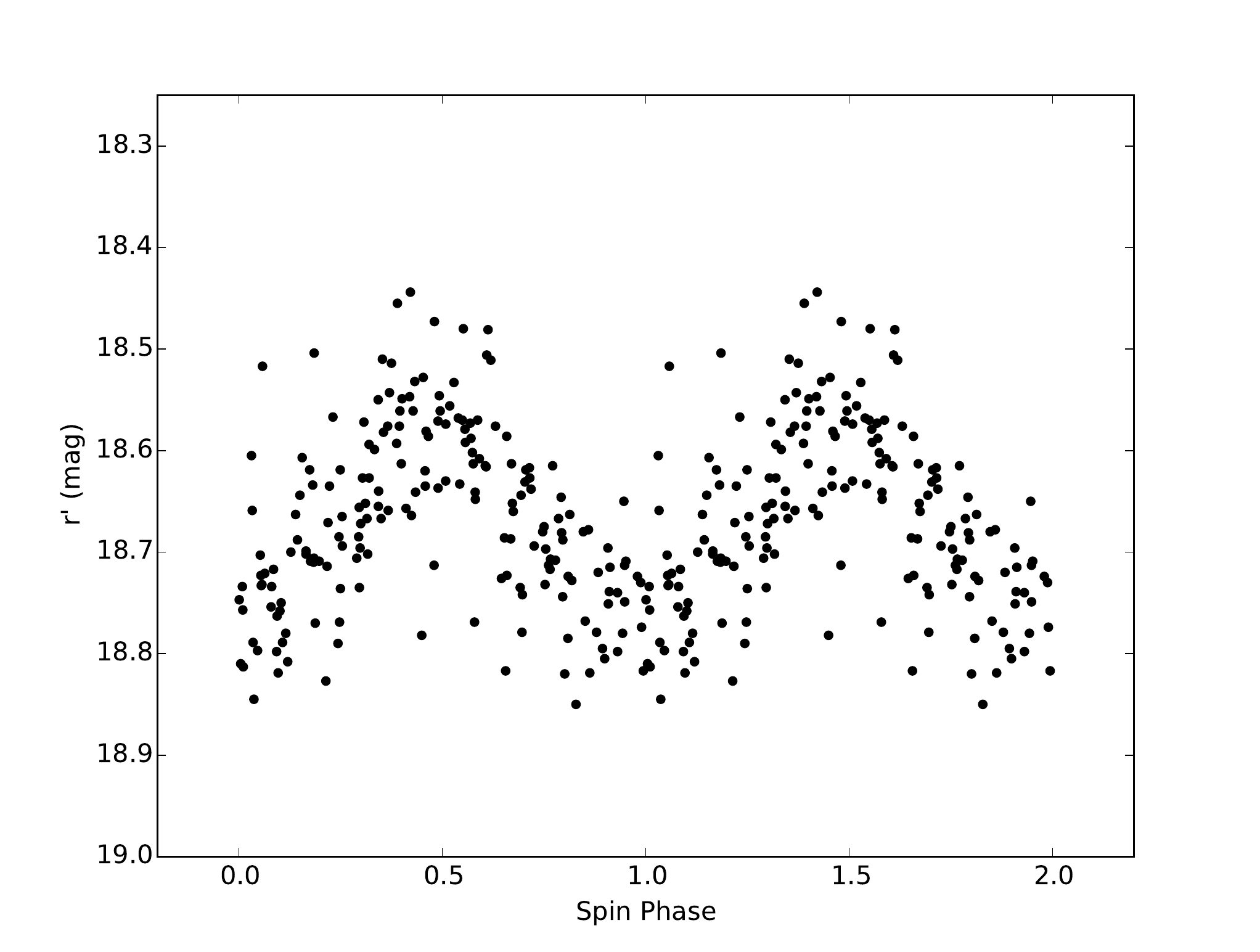}
\caption{The top panel is the r' DECam light curve (error bars are present but comparable to the marker size). The middle panel shows the data folded on P$_{\rm orb}$ = 0.358 days. The phase-folded light curve is also binned in phase bins of three consecutive data points to show the secondary eclipse more clearly. The bottom panel is the light curve folded on the P$_{\rm spin}$ = 504.4(5) seconds without binning. Note that the eclipses have been removed from the spin-folded light curve in the bottom panel. }
\end{figure}
%\vspace*{-40mm}
%\begin{center}
%\includegraphics[height=2in,width=2.5in]{CX19oglePgram.pdf}
%	\captionof{figure}{The initial Lomb-Scargle periodogram from the OGLE data. The orbital frequency and spin frequency are labeled with several harmonics visible. Note that the maximum peak corresponds to 1/2 the orbital frequency.}
%\end{center}
%\vspace*{-60mm}
\indent CX19 also appears in the OGLE-IV fields as an eclipsing source with a reported period of P$_{\rm rep}$ = 0.3587 days and an average out-of-eclipse I-band magnitude of 17.91. We performed a period search after removing data points associated with what appear to be outbursts (I $\le$ 17.50) on the OGLE-IV data and refined a fundamental orbital frequency of f$_{\rm orb}$ =  2.78781(2) cycles/day and P$_{\rm orb}$ = 0.358704(2) days. This period is consistent with our DECam period and better constrained. We also recovered the WD spin frequency of f$_{\rm spin}$ = 170.3055(5) or P$_{\rm spin}$ = 0.0058718(3) days = 503.32(3) secs. We note that the maximum peak in both periodograms corresponds to the second harmonic of the orbital frequency and is labelled as 2f$_{\rm orb}$ in Figure~2 b). The I-band light curve and phase-folded light curve can be seen in Figure~4. The spin periods recovered (from DECam and OGLE) differ by 1.08 seconds. Within our quoted error bars for DECam, the errors are consistent to 2$\sigma$. Much like the DECam periodogram, we also find a lack of orbital sideband signatures in the OGLE periodogram (the frequency spikes around f$_{\rm spin}$ in Figure~2 b) are aliases). The middle panel of Figure~4 shows a mean out-of-eclipse I-band magnitude of 17.91(5), an eclipse depth of 1.1 mags and and eclipse duration of $\sim$46 mins. The OGLE light curve is remarkably stable over the 3 year baseline and there does not appear to be any evidence of the source entering a low-accretion state. The secondary eclipse is prominent around phase 0.5 with an approximate depth of 0.2 mags. The I-band light curve folded on the spin period is shown in the bottom panel of Figure~4. The spin period is rather stable showing an amplitude of $A$ = 0.17 mags.\\
%\indent As a check on the orbital period determination, we applied a bootstrapping method to the data where we sampled randomly creating a new data set of the OGLE data. We then ran a period search on the ``new'' data set. This was done for 1000 different random data sets concluding with the most frequent orbital period being 0.358704(5) days. 
\indent Due to the smaller uncertainty from the OGLE data, we are inclined to adopt the the fundamental parameters of CX19 as P$_{\rm orb}$ = 0.358704(2) days and P$_{\rm spin}$ = 0.0058718(3) days = 503.32(3) secs.\\ 
\begin{figure}
\centering
\includegraphics[scale=0.43]{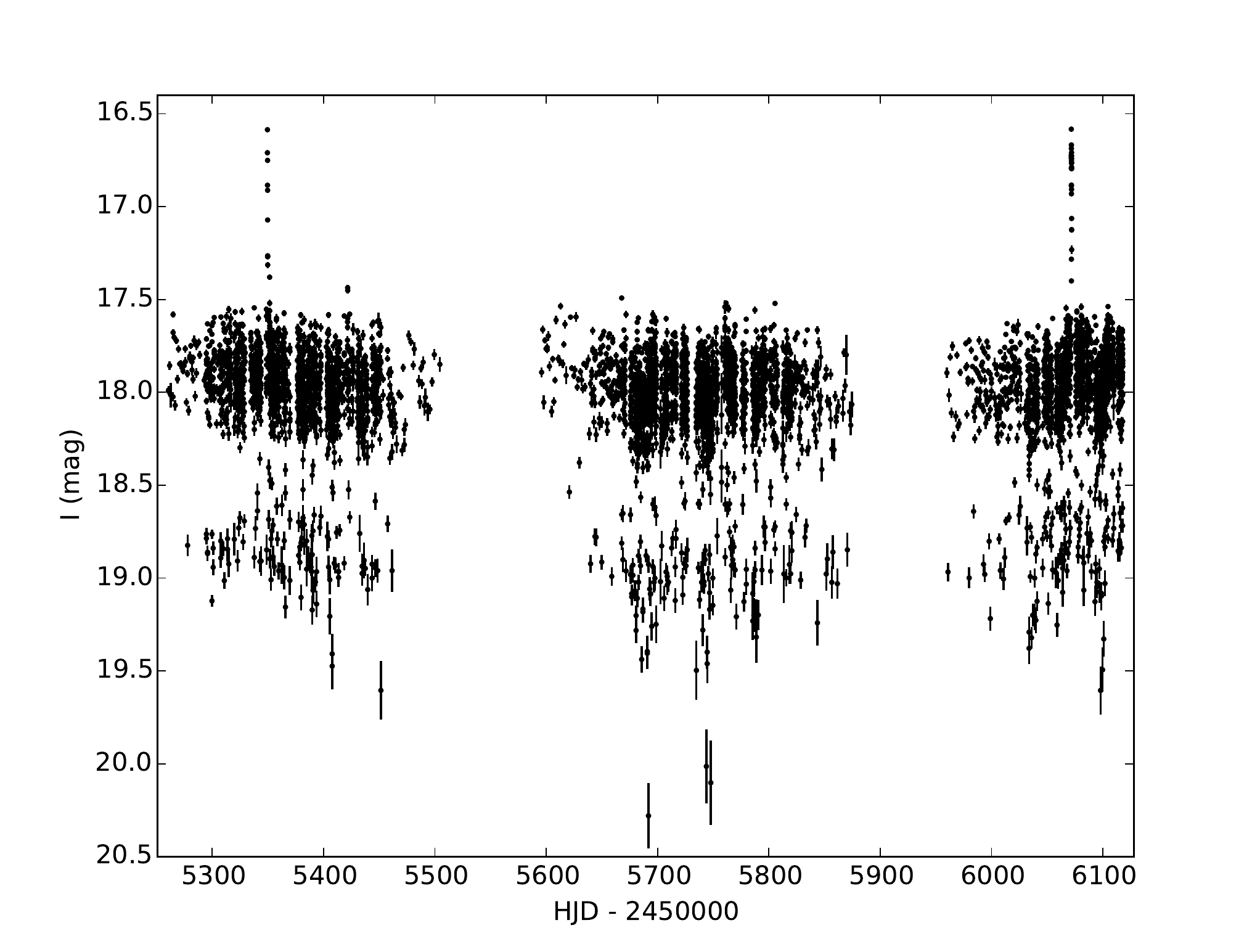}
\includegraphics[scale=0.43]{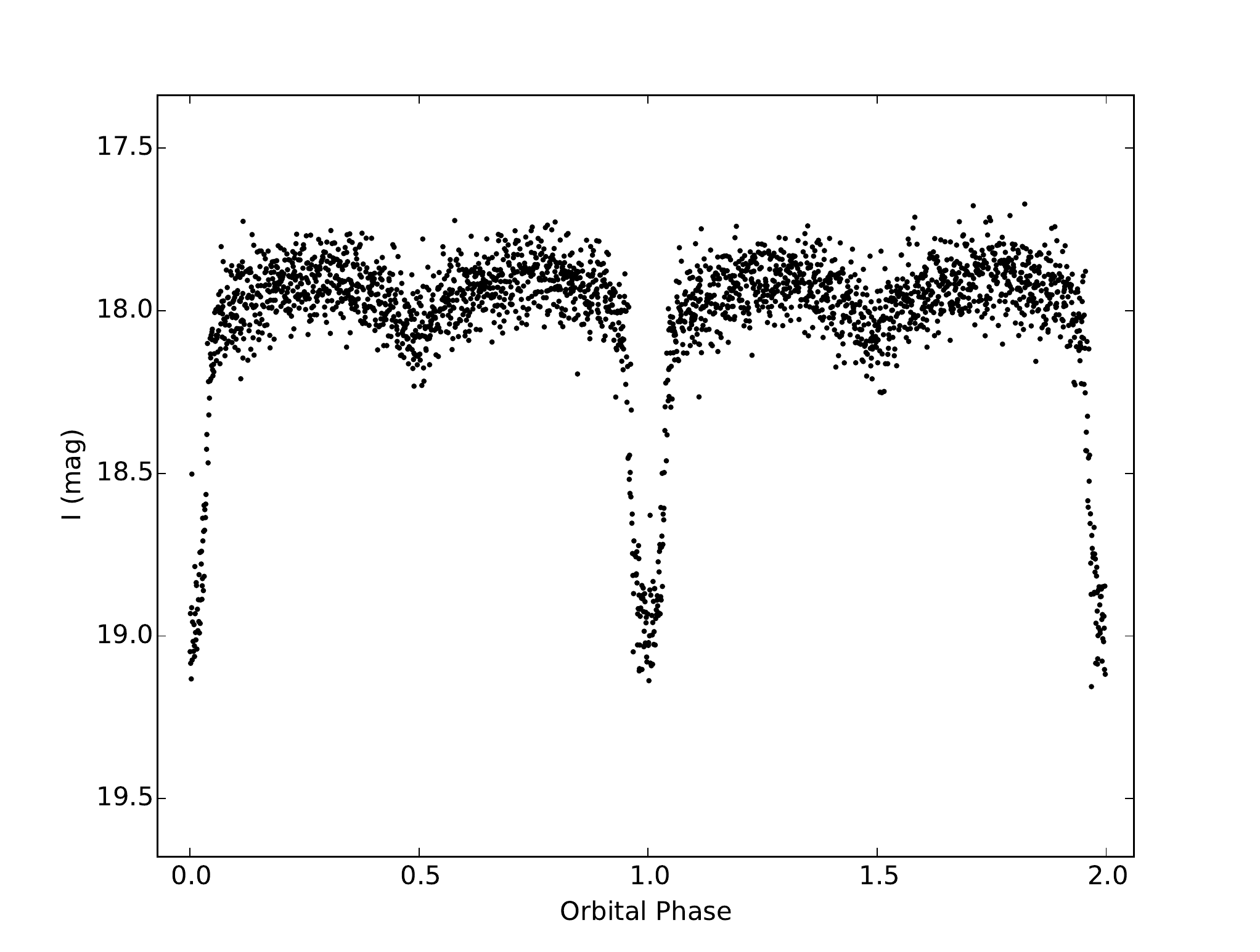}
\includegraphics[scale=0.43]{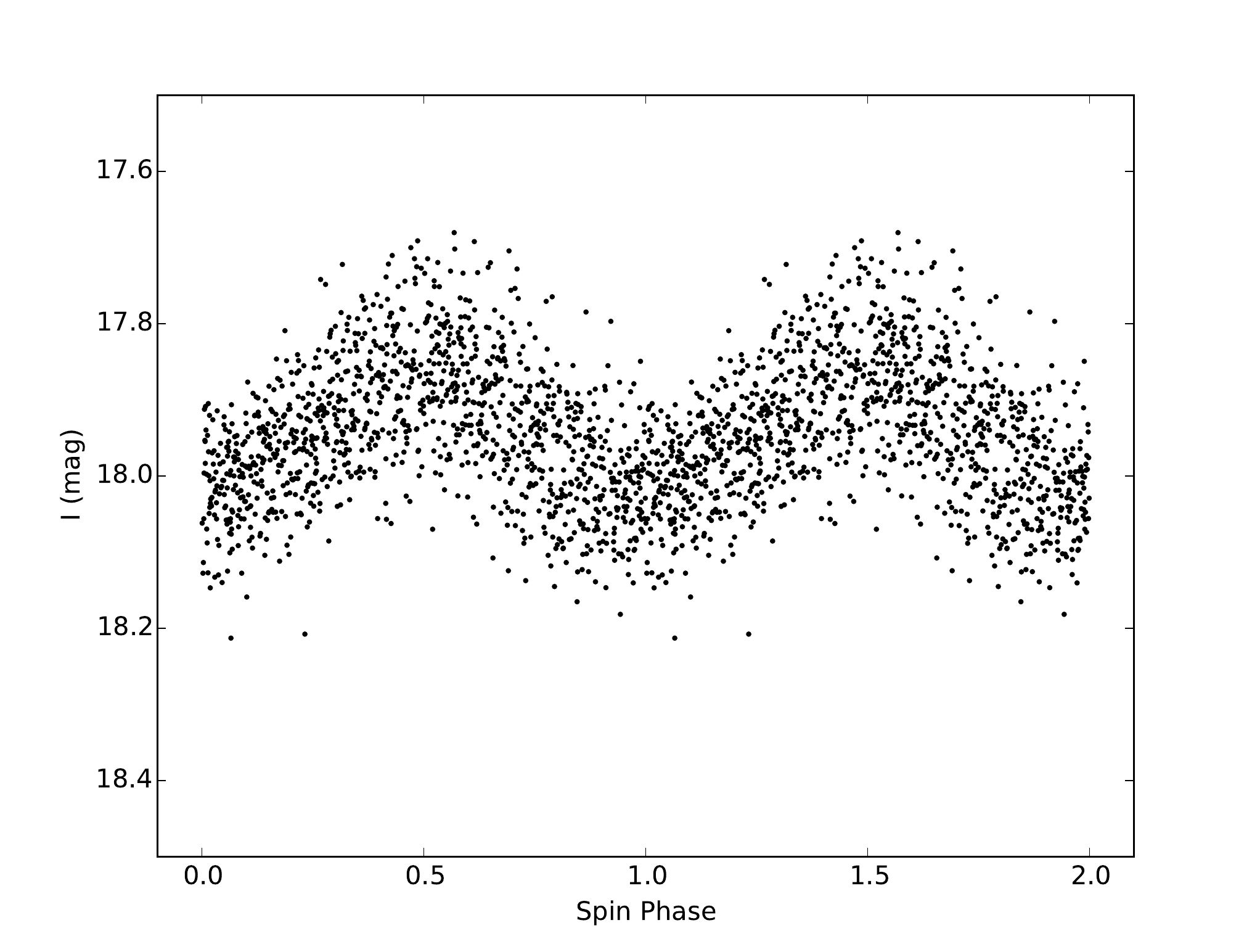}
\caption{The I-band OGLE light curve and the phase-folded light curve using P$_{\rm orb}$ = 0.358704 days can be seen in the top and middle panel, respectively. The bottom panel shows the light curve folded on the spin period of 503.32 secs.}
\end{figure}
\indent With the refined period from the OGLE data, we then used the parabolic fitting technique described above to fit a parabolic function to the phase-folded OGLE data to find the best fit for the epoch, T$_0$, and ephemeris. With a $\chi^2_{red}$ = 3.34 (410 degrees of freedom), we find the best fit to be:
\begin{equation}
{\rm HJD(eclipse)} = 2455691.8581(5) + 0.358704(2)\times {\rm N} 
\end{equation}
%The flickering in both the DECam and OGLE light curves is on the order of $\sim$0.4 mags and disappears during the primary optical eclipse.
%\vspace*{-50mm}

\subsection{Optical Outbursts} Upon inspecting the OGLE data in Figure~4 (top panel), two optical outbursts were discovered and can be seen in a close up view in Figure~5. Since we do not observe the beginning of the outbursts, we can place lower limits on the duration and amplitude and approximate upper limits based on the cadence of the OGLE observations ($\sim$ 1 day). The first outburst occurred in 2010 (HJD = 2455349) (top panel of Figure~5) and the observed peak was at I = 16.58, approximately 1.3 mag brighter than the mean out-of-eclipse brightness of 17.91. From observed peak to the end of the observed decline, the outburst lasts $\sim$6.2 hours. The second outburst occurred in 2012 (HJD = 2456071) (bottom panel of Figure~5) and is similar in observed peak brightness reaching an I magnitude of 16.60, but covering $\sim$8.6 hours, the length of the OGLE observations for that night. Given the ephemeris in equation (2), we find that the outburst profiles have been superimposed on the eclipse events expected to occur at HJD 2455349.6544(5) for the first outburst and HJD 2456071.7256 for the second. The dashed lines in Figure~5 represent where the mid-eclipse times occur. We point out that the eclipse during the 2012 outburst is visible in Figure~5, but not readily apparent in the 2010 data in Figure~5.\\% A close-up view of the outbursts is shown in the top and bottom panels of Figure~3.\\ %One can clearly see that no eclipse points are visible during each outburst. This confirms that the outbursts must occur in the accretion disc. \\
\begin{figure}
\centering
\includegraphics[scale=0.5]{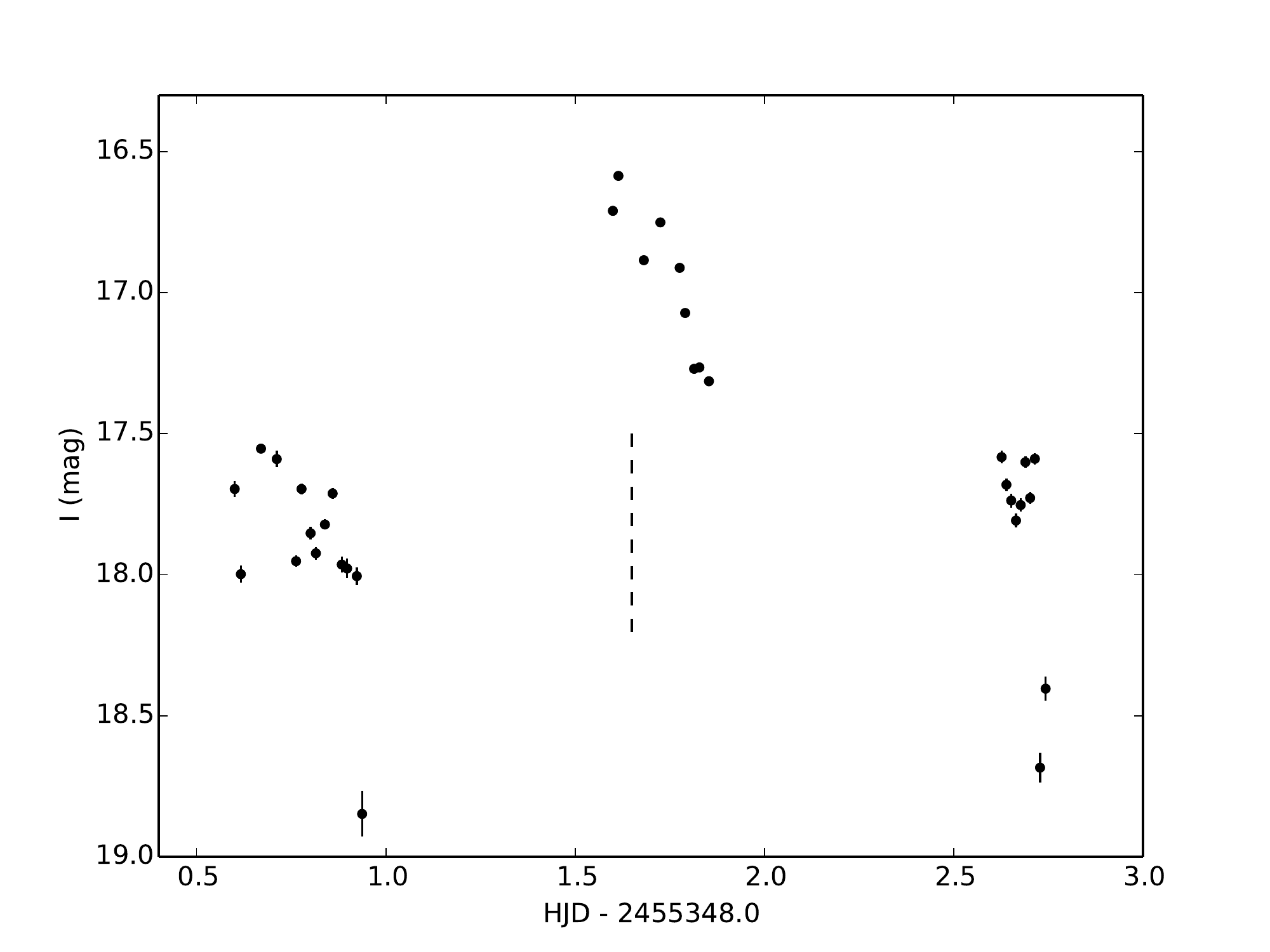}
\includegraphics[scale=0.5]{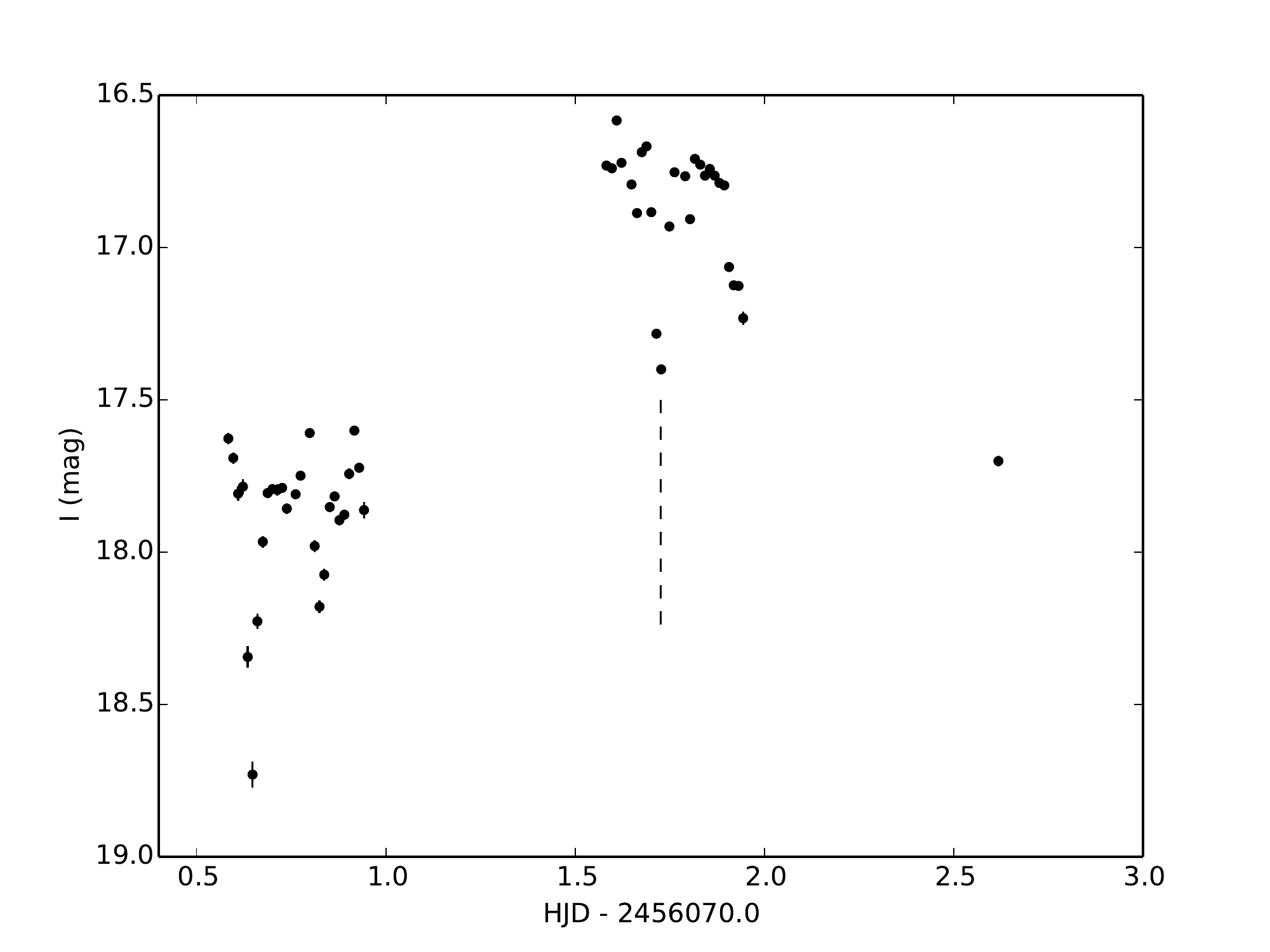}
\caption{The two outbursts seen in the OGLE data in 2010 and 2012 (top and bottom panel, respectively). The dashed lines represent where the middle eclipses occur using equation (2).}
\end{figure}

\subsection{X-ray Features} After reprocessing the CXO data for each of the three OBSIDs (8742, 8753 and 9563), the X-ray data were extracted from all the event files. We extracted a total of 425 counts in the range of 0.3--8.0 keV, the softest being 0.51 keV and the hardest being 7.68 keV. The majority of counts reside in the hard X-ray regime $>$4.0 keV. We define the hardness ratio as HR= (H-S)/(H+S), where S and H are the soft (0.3--2.5 keV) and hard (2.5--8 keV) energy band count rates \citep{Jonker2011}. We find a hardness ratio of HR = 0.70 $\pm$ 0.03 derived from the distribution of X-ray counts making CX19 a hard X-ray source. The X-ray light curve and eclipse is presented in Figure~6. The eclipse was detected in the middle of the 15 ks exposure (OBSID 9563) and appears to be a total eclipse lasting $\sim$31 mins. The X-ray light curve shows variability by a factor of $\sim$2 on timescales of several minutes outside of the eclipse although we do not detect any significant spin modulation in the data.\\ 
\begin{figure}
\centering
\includegraphics[scale=0.5]{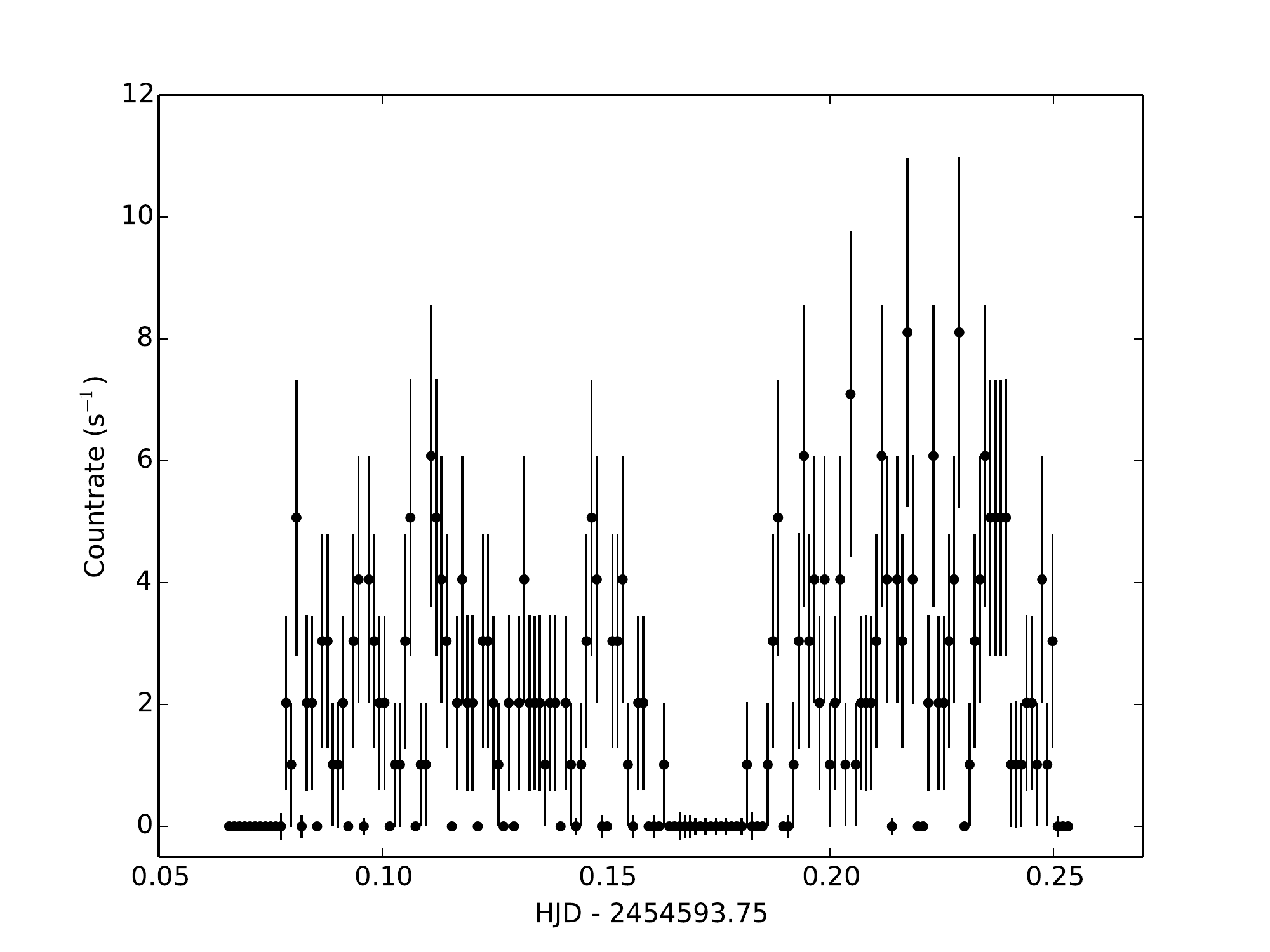}
\caption{The 0.3--8.0 keV Chandra X-ray light curve of CX19 (in bins of 100 sec) showing the X-ray eclipse.}
\end{figure}
We then used the specextract CIAO tool to extract spectra and backgrounds, and create response files, for all three observations. We grouped the longer ObsID 9563 spectrum by 20 counts/bin, and combined the ObsID 8742 and 8753 spectra into one spectrum, which we binned by 10 counts/bin. We fit both spectra with simple models in XSPEC, starting with an absorbed power-law (using abundances from \citealt{Wilms2001}). The power-law fit, forcing both spectra to have the same parameters, is poor ($\chi^2_{red}$ =  1.89 for 16 degrees of freedom), and shows an excess of unmodeled flux in the ObsID 9563 spectrum at low energies. Allowing the $N_H$ and normalization to differ between observations gives an acceptable fit ($\chi^2_{red}$=1.05), but the power-law index is then forced to -0.4.\\ 
\indent Considering the identification of this source as a intermediate polar, we tried a partial covering absorber with a high-temperature (fixed to 30 keV) mekal (thermal plasma, bremsstrahlung dominated; e.g. \citealt{Liedahl1995}) model, as has typically been used to describe intermediate polars (e.g. \citealt{Norton1989}). Thus, the model is tbabs$\times$pcfabs$\times$mekal. We tie the Galactic absorption (tbabs) and intrinsic absorption (pcfabs) between the spectra, but permit the covering fraction and mekal normalization to be free between the spectra. This gives an excellent fit (reduced $\chi^2_{red}$=0.84 for 14 degrees of freedom), and is shown in Figure ~7.  In this fit, the Galactic absorption is $1.7^{+1.2}_{-0.9}\times10^{22}$ cm$^{-2}$, the partial absorber is $23^{+12}_{-7}\times10^{22}$ cm$^{-2}$, the covering fractions are 91$^{+4}_{-6}$\% and 79$^{+11}_{-21}$\% (for the longer and shorter observations respectively). The mekal temperature is $>$13 keV; we fixed it to 30 keV. The implied intrinsic 0.5-10 keV fluxes are $3\times10^{-12}$ and $4\times10^{-12}$ ergs cm$^{-2}$ s$^{-1}$, for the longer and shorter observations respectively. (This includes correction of the longer observationÕs flux for the $\sim$20\% of the exposure during eclipse.) At the minimum distance of 2.1 kpc (see Section 3.6), this converts to a 0.5-10 keV luminosity of 1.5--2.0$\times10^{33}$ ergs s$^{-1}$ between the two quoted X-ray luminosities making up the range. This is somewhat high for CVs, but not unprecedented for magnetic CVs, especially those that are X-ray selected (e.g. \citealt{Beuermann2004}, \citealt{Pretorius2014}). We would like to mention that we performed similar fits on the unbanned data using a C-stat X-ray spectral fitting approach and found essentially equivalent results. 

\begin{figure}
%\begin{\raggedleft}
\hspace*{32mm}
%\raggedleft
\includegraphics[height=3.5in,width=3.0in,angle=270]{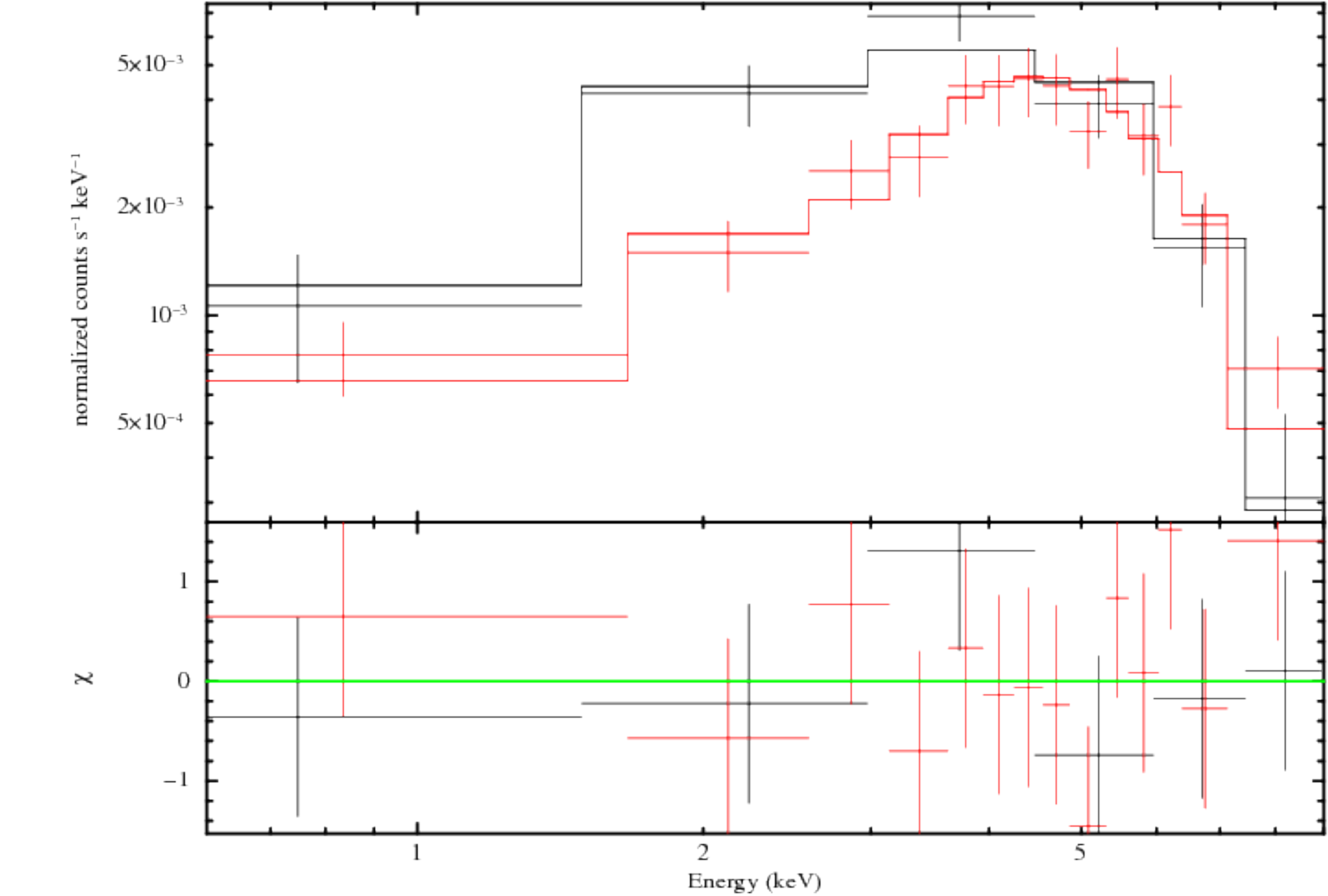}
\caption{X-ray spectral energy distribution and best fit. The top panel shows two spectral fits: OBSID 9563 (red) and the combined spectra of OBSIDs 8742 and 8753 (black). The bottom panel shows the residuals from each fit.}
%\end{\raggedleft}
\end{figure}
%We also find that the source spectrum can be described by a power law with photon index of 2, a Galactic H{\sc i} column density of N$_H$ = 1.1 $\times$ 10$^{22}$ cm$^{-2}$ and count rate of 0.033 counts s$^{-1}$. This provides a good fit at an absorbed/unabsorbed 0.3--8.0 keV flux of 5.0 $\times$ 10$^{-13}$/1.1 $\times$ 10$^{-12}$ erg cm$^{-2}$ s$^{-1}$, respectively. Using the minimum distance of $d \approx$ 2.1 kpc, this converts to a 0.3--8.0 keV luminosity upper limit of 5.5 $\times$ 10$^{32}$ erg s$^{-1}$. Such an X-ray luminosity is consistent with a white dwarf accretor in a CV, especially IPs \citep{jon11}.
\subsection{Spectroscopic Features} The averaged VIMOS spectrum (Figure~8, top panel) shows H$\alpha$ in emission with a broad (intrinsic FWHM of $850 \pm 20 $ km s$^{-1}$) single-peaked line profile with EW = $ 23.2 \pm 0.2$ \AA. He{\sc i} $\lambda\lambda5876, 6678$ are also present in emission with EW $\sim 3$ \AA. At redder wavelengths, CX19 shows the emission lines P12/11/10/9 $\lambda8750/8863/9015/9229$ of the Paschen series (with P$n$ standing for the Paschen $n - 3$ transition). The Paschen line profiles are broad ($\sim1400$ km s$^{-1}$ FWHM)  with P9 being the strongest of them. DIBs at $\lambda5780$ and $\lambda6284$ appear to be present, but affected by residual emission from the sky background. In addition, the interstellar NaD doublet is saturated. Absorption lines from the donor star are not detected. In particular, there is no definitive evidence of the Ca{\sc ii} triplet in absorption (or emission). Finally, to remark here, the VIMOS observation covered orbital phases $0.89-0.95$.
%The averaged VIMOS spectrum of CX19 (Figure~8, top panel) shows H$\alpha$ in emission with a broad and single-peaked line profile with intrinsic FWHM of $850 \pm 20 $ km s$^{-1}$ and EW = $ 23.2 \pm 0.2$ \AA. He{\sc i} $\lambda5876, 6678$ are also in emission, but weaker (EW $\sim 3$ \AA). At longer wavelengths, CX19 shows the emission lines P12/11/10/9 $\lambda8750/8863/9015/9229$ of the Paschen series where P$n$ stands for the Paschen $n - 3$ transition. The Paschen lines are broad ($\sim1400$ km s$^{-1}$ FWHM)  with P9 being the strongest of them. In addition to the saturated interstellar NaD doublet, DIBs at $\lambda5780$ and $\lambda6284$ appear to be present, but affected by residual emission from the sky background.  Photospheric lines from the donor star are not detected, in particular, there is no definitive evidence of the Ca{\sc ii} triplet in absorption or emission. The VIMOS observation covered orbital phases $0.89-0.95$. 
\begin{figure}
\centering
\includegraphics[scale=0.5]{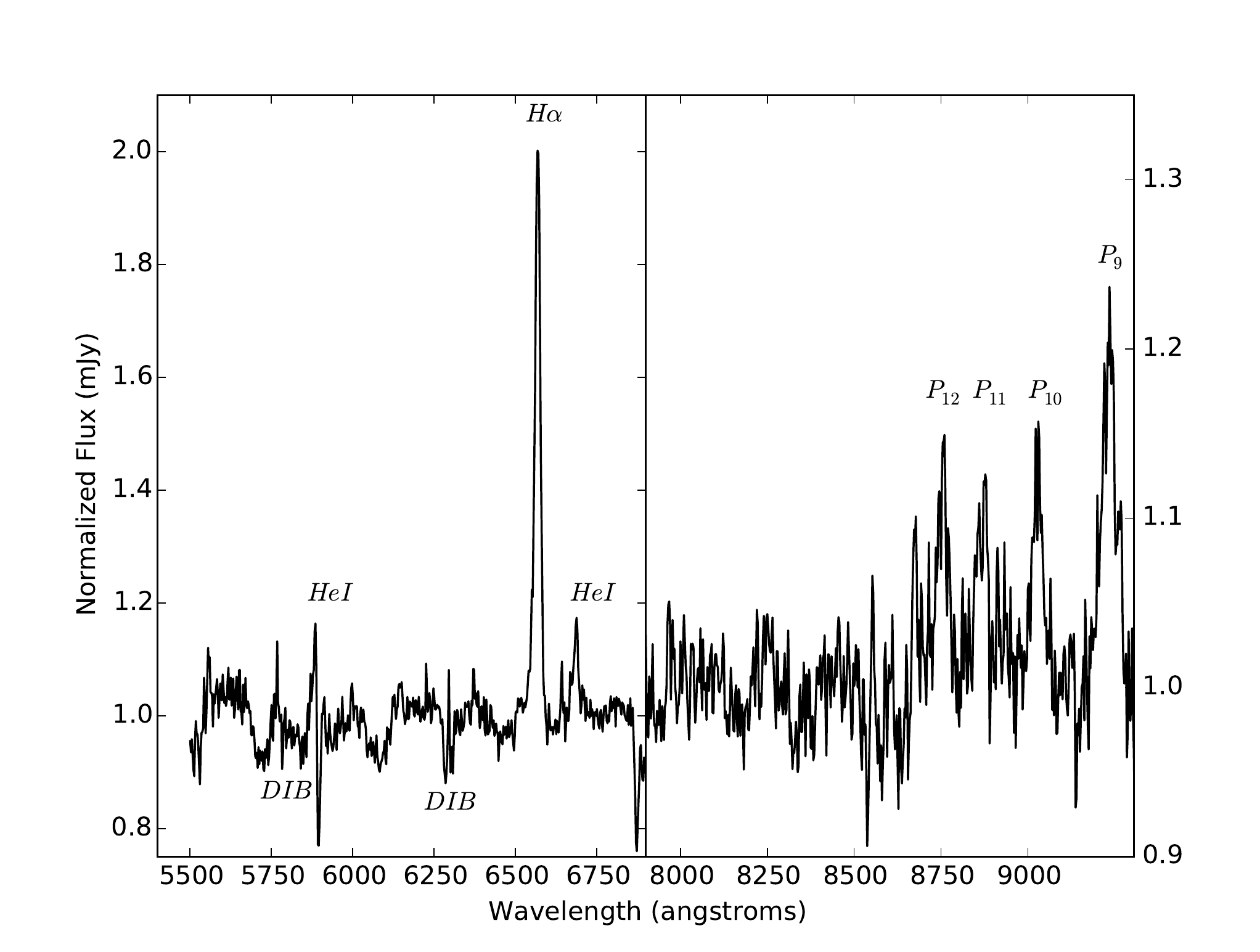}
\includegraphics[scale=0.5]{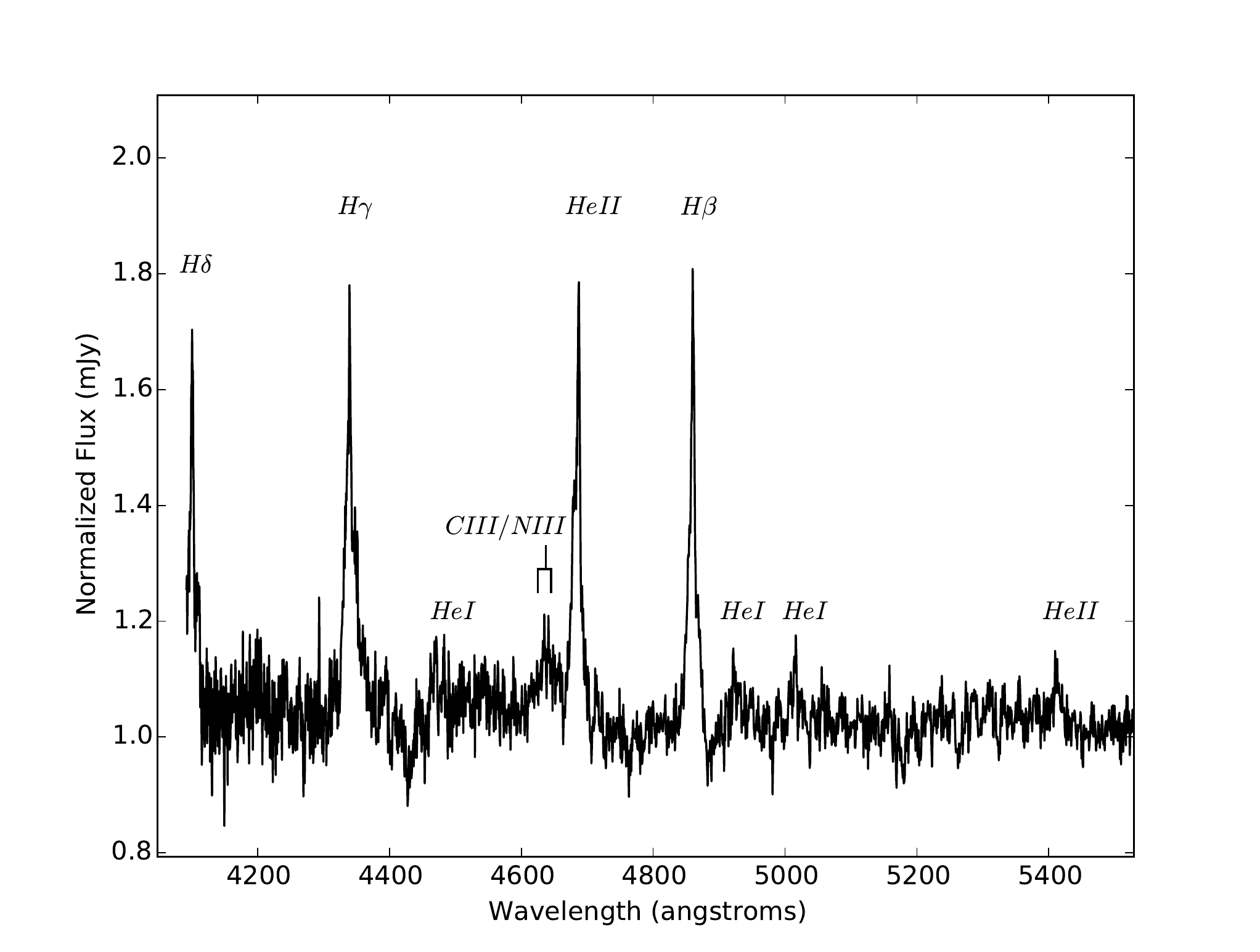}
\caption{Normalized and averaged spectra of CX19 obtained with VIMOS (top panel) and FORS2 (bottom panel). The most prominent features are labeled.}
\end{figure}
The bottom panel of Figure~8 shows the averaged FORS2 observations. The blue coverage and spectral resolution provided by the data allows us to detect and characterize emission lines from H$\gamma$, H$\beta$ and He{\sc ii} $\lambda4686$. Their line profiles are composed at least of a broad and a red-shifted narrow component. From a 1D-Gaussian fit, we derive the lines' Heliocentric radial velocities and FWHMs given in Table~1. This table also provides equivalent widths (EWs) and the line parameters for the H$\alpha$ line in the VIMOS observation. The FWHMs are corrected for the instrumental resolution. The H$\beta$ and He{\sc ii} lines have similar broadening and strength (840-940 km s$^{-1}$ FWHM, $\sim10$ \AA~EW). Weaker C{\sc iii}/N{\sc iii} Bowen blend emission near $\lambda4640$ and {He{\sc ii} $\lambda 5412$ are also present. The narrow emission line component from He{\sc i} $\lambda\lambda4912,5016$ is also visible while He{\sc i} $\lambda4471$ is not detected. Photospheric features from the WD or donor star are absent except perhaps for a feature at $\lambda5712$ that matches the expected position for Mg{\sc ii}. However it could be an artifact. Other absorption features at $\lambda\lambda4428,4882$ are diffuse interstellar bands. The FORS2 observations covered orbital phases 0.01-0.84 with a phase resolution of 0.014 although this was not continuous coverage.\\ 
\indent Figure~9 shows a zoom-in of the evolution of the line profile for the He{\sc ii} line and the H$\beta$ line. The emission lines are stronger at mid-eclipse than outside eclipse, indicating that the source of their emission is less concentrated towards the orbital plane than the source of the continuum. The peak of the $\lambda4686$ appears to be shifted to the blue at phase 0.58 and 0.62 and close to the rest wavelength at other phases. The H$\beta$ is shifted to the red in Figure~9 (phases 0.62-0.84) and behaves opposite to the He{\sc ii} line which could be consistent with tracing out the hot spot of the disc. We do point out that double-peaked emission lines at eclipse phases suffer a Z-wave disturbance which results from the emitting region being eclipsed. In this type of scenario, the blue-shifted disc emission is first eclipsed followed by the eclipse of the red-shifted emission. The donor star usually occults the disc regions moving towards the observer and subsequently blocks the receding regions from view of the observer. We do not see a clear double-peaked emission line profile, but we are not ruling the possibility of one out and note that there is not a clear picture of the geometry of CX19.

\begin{figure}
\centering
\includegraphics[scale=0.65]{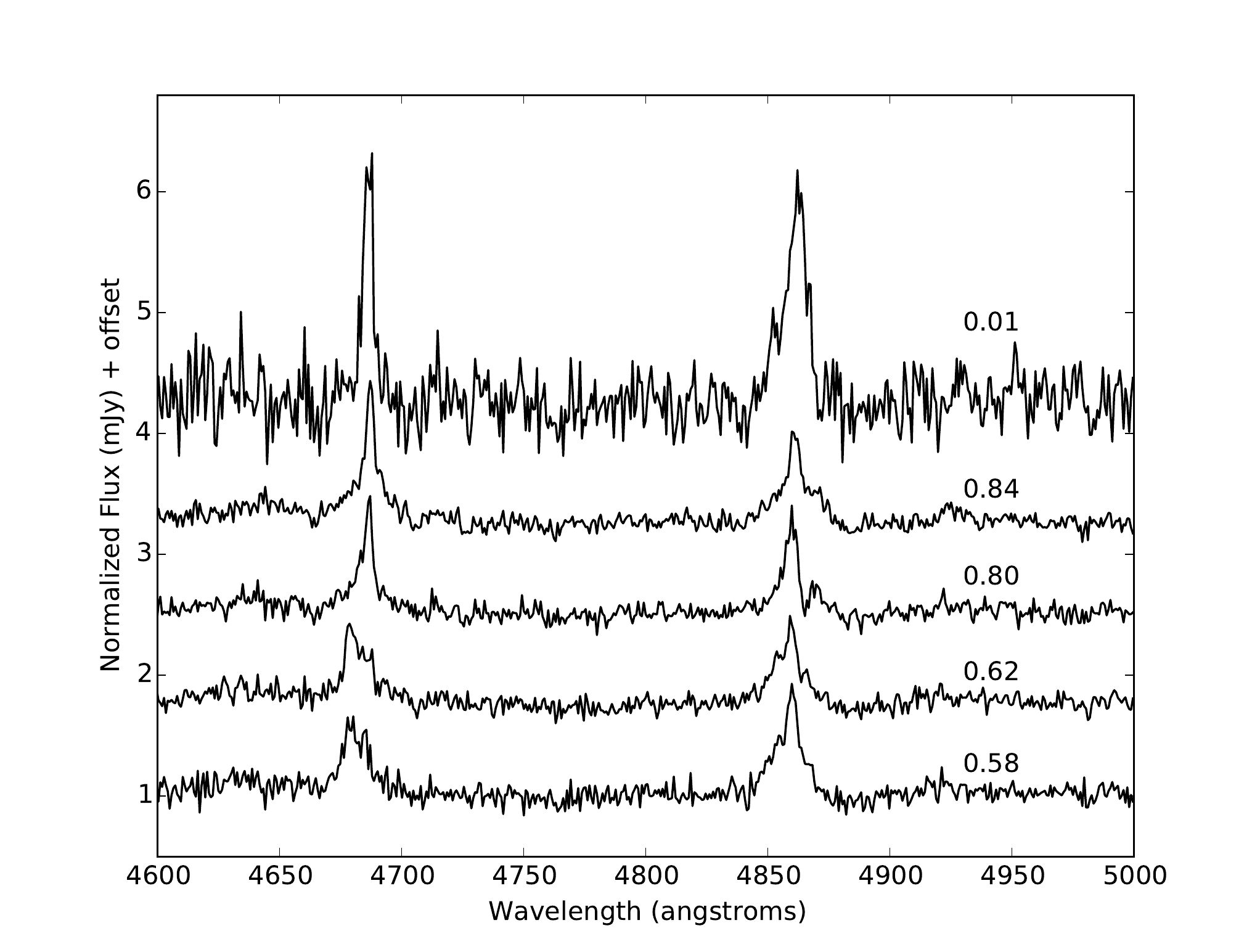}
\caption{A zoom-in of the FORS2 spectra and the evolution of the line profile for the He{\sc ii} line and the H$\beta$ line. Each FORS2 spectrum is labelled corresponding to the orbital phases 0.58, 0.62, 0.80, 0.84 and 0.01. }
\end{figure}

\begin{table}
\caption{Spectroscopic measurements of the strongest emission lines}
\label{log}
\begin{center}
\begin{tabular}{lcccc}
\hline
          & & RV    &EW  &  FWHM        \\
Line   & &  (km/s)       & (\AA)  & (\AA)      \\ 
\hline
\hline
H$\gamma$                      & &    $  -20 \pm 20   $ &  $ 13.6 \pm 0.3 $ & $ 20.7  \pm 0.8  $                    \\    
He{\sc ii} $\lambda4686$ & &    $ -80  \pm 20  $ &   $10.2   \pm 0.2 $ & $ 14.6 \pm 0.6 $            \\      
H$\beta$                           & &    $-110 \pm 10 $  &  $ 9.8 \pm 0.2    $ &  $ 13.6 \pm 0.4  $            \\  
H$\alpha$                         & &     $  230 \pm  8  $ &   $ 23.2 \pm 0.2 $  &  $ 18.0  \pm 0.4 $              \\      
\hline
\hline
\end{tabular}		      
\end{center}
\end{table}

\subsection{Distance Estimate} To calculate an approximate distance, we began by estimating the spectral type of the donor star from the density-period relation for Roche lobe overflowing stars given by:
\begin{equation}
\rho \cong 110\times {\rm P}^{-2}_{\rm orb, hr} \hspace{4pt} {\rm g} \hspace{2pt} {\rm cm}^{-3}
\end{equation}
where the orbital period is given in hours \citep{Frank2002}. We find an approximate density of $\rho \approx$ 1.5 ${\rm g} \hspace{2pt} {\rm cm}^{-3}$ which is consistent with a G3V-G5V spectral type donor using densities from \cite{Allen1973}. In our calculation for the distance estimate, we will assume a main sequence companion. We do point out that the companion is more likely a less dense, later-type star which is usually seen in longer period systems like CX19 and that the following discussion is only a best estimate based on the data at hand. In an attempt to estimate the reddening, we determine the observed (R--I) colour of CX19 using the I-band magnitude (in eclipse) of I = 19.05 $\pm$ 0.01 and the transformation equations between Johnson-Cousins and SDSS u'g'r'i'z' magnitudes derived by \cite{Lupton2001}. The observed colour (R--I) $\approx$ -0.13 is much bluer than the intrinsic colour of known G3V-G5V stars ($\approx$ 0.35--0.36) compiled in Eric Mamajek's online stellar tables\footnote[3]{http://www.pas.rochester.edu/$\sim$emamajek/}. This is expected since the spectrum is disc dominated without any apparent absorption features from the donor star. Also, the emission line features appear to get stronger during mid-eclipse (Figure~9) indicating that excess blue light is present and that the disc is not {\it completely} eclipsed. We estimate the maximum extinction to the source by using the ``Bulge Extinction and Metallicity Calculator" (BEAM)\footnote[4]{http://mill.astro.puc.cl/BEAM/calculator.php} online tool which gives A$_{K}$ $\approx$ 0.7180 for a radius of 1.1' centered at the source coordinates of CX19. Since the G-type star does not have to be an ordinary main sequence star, especially if it is in an interacting binary system, we can calculate a distance estimate based on the following assumptions concerning extinction and absolute magnitude. Using A$_{K}$ = 0.225$\times$A$_{I}$ \citep{Allen1973}, the absolute magnitude of a G5 type star (M$_I$ = 4.242), and the apparent magnitude (I = 19.05), we find a minimum distance of $d \approx$ 2.1 kpc. The caveat to this conclusion is that if we have a later spectral type for the companion, the surface brightness would be lower and the distance would be even closer. 

\section{Discussion} There are several main observational characteristics of IPs that define the subclass. In short, IPs are characterized by strong, hard X-ray emission, high-excitation spectra, and stable optical and X-ray pulsations in their light curves. From an optical standpoint, the system should show a stable photometric period which can be determined from the spectroscopic radial velocity curve and/or the photometric modulation. Due to the radial accretion from the magnetic field lines on the poles of the WD surface, X-ray pulsations or eclipses at the same, or very similar, P$_{\rm opt}$ should be present. Depending on the geometry of each system and the inclination angle, X-ray eclipses can be full or partial (grazing). To complete the list for inclusion in to the IP subclass from \cite{Patterson1994}, there should be pulsations in the He{\sc ii} emission lines, which almost certainly arise from photoionization by the central X-ray source, circular polarization present, the existence of ``beat" or ``sideband" frequencies in optical and X-ray light, usually on the low-frequency side of the main signal and a very hard X-ray spectrum, often with a strong signature of low-energy absorption. These properties are shared among all confirmed IPs, but every property is not required for a system to be labeled as an IP.\\
\indent At the time of writing, the list of confirmed IPs composed by K. Mukai\footnote[5]{http://asd.gsfc.nasa.gov/Koji.Mukai/iphome/iphome.html} contains 47 systems. Among the confirmed IPs, the orbital period range spans from about 0.05--2.0 days. The WD spin period ranges from $\sim$33--4000 seconds. Mukai's IP list contains seven systems that show confirmed eclipses, five of which are grazing/partial: FO Aqr, TV Col, EX Hya, BG CMi and V598 Peg. The other two, DQ Her (P$_{\rm orb}$ = 4.646 hr) and XY Ari (P$_{\rm orb}$ = 6.064 hr), are deeply eclipsing IPs. It has been shown previously that there appears to be a loose relation between the two periods inherent in most IP systems, mainly P$_{\rm spin}$/P$_{\rm orb}$ $\sim$ 0.1 for systems with P$_{\rm orb}$$<$ 5.0 hours. The most extreme systems in the list have P$_{\rm spin}$/P$_{\rm orb}$ $<$ 0.01 and are found at very long orbital periods, i.e. GK Per (P$_{\rm orb}$ = 47.9 hrs, P$_{\rm spin}$ = 351.3 secs). CX19 has a P$_{\rm spin}$/P$_{\rm orb}$ = 0.016(6)  as would be expected for a long orbital period IP.\\
\indent \cite{Norton2004} showed that there exists a relation for a large range of P$_{\rm spin}$/P$_{\rm orb}$ and WD magnetic moments, $\mu_{\rm WD}$, for magnetic CVs. This relation is illustrated in Fig.~2 of \cite{Norton2004} for an assumed mass ratio of $q$ = M$_2$/M$_1$, where M$_2$ and M$_1$ are the donor star and WD masses. As an estimate for the magnetic moment of CX19, we use a P$_{\rm orb}$ = 0.358704 days or $\sim$ 8 h and P$_{\rm spin}$/P$_{\rm orb}$ $\sim$ 0.02 in their Figure~2 to find a very conservative magnetic moment of the WD of $\mu_{WD}$ $\sim$1--2 $\times$ 10$^{33}$ G cm$^3$. For a longer period IP, this is a relatively high magnetic moment and the system is most likely going to evolve into a synchronous polar.\\
\indent We classify CX19 as a deeply-eclipsing IP system with an average magnitude depth of $\Delta$ mag $\sim$ 1.1 in the I-band. \cite{Warner2009} identified V597 Pup as the third deeply eclipsing IP and \cite{Aungwerojwit2012} presented the fourth deeply eclipsing IP, IPHAS J062746.41+014811.3, with a P$_{\rm spin}$/P$_{\rm orb}$ = 0.075 and an eclipse depth of $\Delta$ mag $\sim$1.5. Recently, \cite{Esposito2015} have classified Sw J2014 as a deeply eclipsing IP with an orbital period of 3.44 hrs. CX19 is the sixth system to exhibit deep optical (almost flat-bottomed) eclipses. The similar strengths and FWHMs observed for the Balmer and He{\sc ii} $\lambda4686$ lines in CX19 resemble the emission line properties of the 9.81 hr orbital period intermediate polar RXS J154814.5-452845, although in this system, both WD and donor star have been detected in the optical spectra \citep{Demartino2006}.  Using the relation from \cite{Horne1985} we estimate the inclination angle of the system using the eclipse width duration of 6\% and a conservative mass ratio of $\sim$1.  We find an approximate inclination angle of $\sim$73$^{\circ}$.\\ 
\indent Only one other IP, XY Ari, has been confirmed to have both deep eclipses in the X-ray {\it and} another bandpass, the IR \citep{Hellier1997}. XY Ari shows X-ray eclipses recurring with a 6.06 h orbital period but since it is hidden behind the the molecular cloud Lynds 1457, the optical flux is extinguished to V$>$23. The X-ray light curve of XY Ari also exhibits a clear, pulsed 206 s spin period out-of eclipse. CX19 shows an X-ray eclipse in Figure~6 with a flat bottom lasting for $\sim$0.06 of the phase coverage or $\sim$31 min, with no clear modulation out-of-eclipse. To our knowledge, CX19 is the first X-ray selected IP that shows clear, deep eclipses in both the optical and X-ray. DQ Her does show a partial X-ray eclipse, but due to the high inclination ($\sim$87$^{\circ}$) of the system, most X-rays are obscured by the accretion disc \citep{Mukai2003}.\\
\indent Optical outbursts such as those seen in Figure~5 have also been observed in several magnetic CV systems, most notably in the IPs TV Col and V1223 Sgr \citep{Van1989}. V1223 Sgr has shown outbursts lasting $<$1 day and $>$1 mag-amplitude, similar to the CX19 outbursts. In the case of EX Hya, outbursts have been observed to reach 4 magnitudes in amplitude lasting for $\sim$2 days. It is thought that these outbursts are a consequence of {\it either} an increased mass transfer rate from the secondary star {\it or} by an instability in the accretion disc itself \citep{Angelini1989}. The idea behind the short outbursts in IPs is explained in terms of the magnetosphere radius. As the temperature in the outer disc region rises,  the high-temperature region gradually starts moving inwards. The result being that the optical brightness increases to maximum rather quickly before the ultraviolet brightness maximum and the duration of maximum is shortened by the inner disc radius, which, in turn, is confined by the magnetosphere radius. Similar to CX19, three outbursts were seen in TV Col lasting $\sim$6 hours \citep{Szkody1984}. \citet{Angelini1989} offers the brief explanation of a small disc around a large magnetosphere to account for the short outburst duration.\\ 

\section{Conclusion} %While analyzing the light curves and spectroscopy of the Galactic Bulge Survey X-ray source CX19, it was apparent that we had uncovered an interacting binary showing many of the same characteristics of intermediate polars. 
CX19 is a deeply-eclipsing system with a period of P$_{\rm orb}$ = 0.358704(2) days along with a modulation on the spin period of P$_{\rm spin}$ = 503.32(3) secs. The ratio of the spin period to the orbital period is P$_{\rm spin}$/P$_{\rm orb}$ = 0.016(6) and is further evidence to support the idea that CX19 is indeed an IP. An X-ray analysis has also confirmed the existence of a total X-ray eclipse with some variability out-of-eclipse in the 0.3--8.0 keV range. Similar to the outbursts seen in V1223 Sgr and TV Col, we have shown that CX19 underwent two clear $\sim$1 mag outbursts lasting between 6--8 hours during the times of mid-eclipse. %The existence of the orbital and spin periods, the hard X-ray emission, optical spectroscopic emission and the outbursts provide evidence that CX19 is a prime candidate IP and should be followed up with time-series analysis in both the X-ray and optical spectroscopic regimes. %If confirmed, CX19 will be the fifth deeply eclipsing IP with an orbital period of P$_{\rm orb}$=0.358705(5) days. 

\section*{Acknowledgments}
This research has made use of the SIMBAD database, operated at CDS, Strasbourg, France, and NASAÕs Astrophysics Data System. CH is supported by an NSERC Discovery Grant. 
Facilities: OGLE, CTIO, APASS DR9, CXO, VLT, VIMOS, FORS2.

\bibliography{biblio1}

\end{document}